\documentclass[aps,prd,superscriptaddress,preprintnumbers,nofootinbib,10pt]{revtex4-2}
\usepackage{multirow}
\usepackage{amsmath}
\usepackage{amssymb}
\usepackage[dvipdf,dvips]{graphicx}
\usepackage{color}
\usepackage{hyperref}
\usepackage{url}
\usepackage{slashed}
\usepackage{subfig}
\usepackage[usenames,dvipsnames]{xcolor}
\usepackage{amsmath}
\usepackage{amsfonts}
\usepackage{float}
\usepackage{amssymb}
\usepackage{epsfig}
\usepackage{graphics}
\usepackage{euscript}
\usepackage{slashed}
\usepackage{epstopdf}
\usepackage[utf8]{inputenc}
\allowdisplaybreaks
\usepackage[normalem]{ulem}
\usepackage{pifont}
\usepackage{dsfont}
\usepackage{graphicx}
\usepackage{latexsym,braket}
\usepackage{tikz-feynman}
\usepackage{tikz-cd}
\usepackage{easyReview}
\usepackage{cancel}
\usepackage[normalem]{ulem}
\usepackage{svg}
\usepackage{cleveref,fouriernc}

\newcommand{\p}{\partial}

\makeatletter
\DeclareRobustCommand{\cev}[1]{%
	{\mathpalette\do@cev{#1}}%
}
\newcommand{\do@cev}[2]{%
	\vbox{\offinterlineskip
		\sbox\z@{$\m@th#1 x$}%
		\ialign{##\cr
			\hidewidth\reflectbox{$\m@th#1\vec{}\mkern4mu$}\hidewidth\cr
			\noalign{\kern-\ht\z@}
			$\m@th#1#2$\cr
		}%
	}%
}
\makeatother

\makeatletter
\newlength{\negph@wd}
\DeclareRobustCommand{\negphantom}[1]{%
  \ifmmode
    \mathpalette\negph@math{#1}%
  \else
    \negph@do{#1}%
  \fi
}
\newcommand{\negph@math}[2]{\negph@do{$\m@th#1#2$}}
\newcommand{\negph@do}[1]{%
  \settowidth{\negph@wd}{#1}%
  \hspace*{-\negph@wd}%
}
\makeatother

\begin{document}

\title{Non-perturbative corrections to the Casimir energy for a scalar field theory with non-linear boundary conditions}

\author{Fabrizio Canfora}
\email{fabrizio.canfora@uss.cl}
\affiliation{Centro de Estudios Cient\'{\i}ficos (CECs), Casilla 1469, Valdivia, Chile}
\affiliation{Universidad San Sebasti\'{a}n, sede Valdivia, General Lagos 1163, Valdivia 5110693, Chile}

\author{David Dudal}
\email{david.dudal@kuleuven.be}
\affiliation{KU Leuven Campus Kortrijk -- Kulak, Department of Physics, Etienne Sabbelaan 53 bus 7657, 8500 Kortrijk, Belgium}

\author{Thomas Oosthuyse}
\email{thomas.oosthuyse@kuleuven.be}
\affiliation{KU Leuven Campus Kortrijk -- Kulak, Department of Physics, Etienne Sabbelaan 53 bus 7657, 8500 Kortrijk, Belgium}

\author{Pablo Pais}
\email{pais@ipnp.troja.mff.cuni.cz}
\affiliation{Instituto de Ciencias F\'isicas y Matem\'aticas, Universidad Austral  de Chile, Casilla 567, 5090000 Valdivia, Chile}
\affiliation{IPNP - Faculty of Mathematics and Physics, Charles University, V Hole\v{s}ovi\v{c}k\'ach 2, 18000 Prague 8, Czech Republic}

\author{Luigi Rosa}
\email{rosa@na.infn.it}
\affiliation{INFN, Sezione di Napoli, Complesso Universitario di Monte S.~Angelo, Via Cintia Edificio 6, 80126 Naples, Italia}
\affiliation{Dipartimento di Matematica e Applicazioni ``R.~Caccioppoli'', Universit\'{a} di Napoli Federico II, Complesso Universitario di Monte S.~Angelo,  Via Cintia Edificio 6,
80126 Naples, Italia}

\author{Sebbe Stouten}
\email{sebbe.stouten@kuleuven.be}
\affiliation{KU Leuven Campus Kortrijk -- Kulak, Department of Physics, Etienne Sabbelaan 53 bus 7657, 8500 Kortrijk, Belgium}

\begin{abstract}
    We consider the case of a free real massive bulk scalar in $D=4$ dimensions, and embed two parallel plates as interfaces on which we impose non-linear boundary conditions, either Dirichlet- or Neumann-like, parameterized by a new coupling constant $g$. This mimics a non-Abelian gauge theory supplemented with boundary conditions on surfaces embedded in the bulk. We present the first evidence for a non-perturbative $\frac{1}{g^2}$ boundary mass generation and its ensuing correction to the standard Casimir energy. This becomes possible by incorporating dynamical corrections to the effective boundary fields, which are used to build in the boundary conditions directly at the action level.
    \end{abstract}
\maketitle
\date{}

\section{Introduction}

Recently, the Casimir effect \cite{Plunien:1986ca,Bordag:2001qi,Milton:2004ya,Bordag:2009zz}, with its origins dating back to \cite{Casimir:1948dh}, saw a revived interest in Yang-Mills gauge theories\footnote{To be more precise, $SU(2)$ and $SU(3)$ gluodynamics, without quark degrees of freedom.}, thanks to non-perturbative lattice QCD Monte-Carlo simulations of $D=(3+1)$ or $D=(2+1)$ setups, with various geometries of interfaces on which standard PEC (perfect color-electric) boundary conditions were applied \cite{Chernodub:2018pmt,Chernodub:2023dok,Ngwenya:2025mpo,Ngwenya:2025cuw}. For simplicity, here we focus on the plate geometry. An interesting feature was observed, namely, the exponential decay in terms of the plate separation. This suggests the emergence of a new, non-perturbative mass scale of a few hundred MeV from the self-interaction of the glue. This mass scale is considerably smaller than the standard lowest mass scale in gluodynamics, the scalar glueball having a mass of around $1.5$~GeV \cite{Athenodorou:2020ani}. In \cite{Chernodub:2023dok}, this new scale was attributed to a novel type of edge (boundary) modes, consisting of a (colorless) bound state of a gluon and its mirror image relative to the plate(s), suggestively named ``glueton'' as analogue of the familiar excitons in solid state physics\footnote{The analogous ``quarkiton'' was studied into more detail in \cite{Chernodub:2025bpz}.}. This does however not answer the question where the mass of these gluetons would come from, as a priori, its constituent gluons are massless. Similar as to plateless QCD (or YM), the pertinent issue still is: how to obtain massive glueballs/gluetons from a priori massless gluons?

Interestingly, in the recent works \cite{Dudal:2025qqr,Karabali:2025olx}, see also \cite{Saharian:2025ioy}, it was shown that very reasonable analytical estimates for the YM Casimir energies/forces of \cite{Chernodub:2018pmt,Chernodub:2023dok,Ngwenya:2025mpo,Ngwenya:2025cuw} were possible to obtain from incorporating a gluon mass scale into the quadratic approximation, either based on $(2+1)D$ works of \cite{Karabali:1995ps,Karabali:2018ael} or the Curci-Ferrari model \cite{Curci:1976bt} in Landau gauge ($(3+1)D$ or $(2+1)D$) that saw a revived interest over the years as an easy to compute with effective description of certain non-perturbative aspects of QCD, see \cite{Pelaez:2021tpq} for an extensive review. The boundary conditions were introduced in those works \cite{Dudal:2025qqr,Karabali:2025olx} via auxiliary fields restricted to the plates, see also \cite{Bordag:1983zk,Golestanian_1998,Dudal:2020yah,Canfora:2022xcx,Oosthuyse:2023mbs,Dudal:2024DEM,Dudal:2024PEMC,Dudal:2024fvo} for more about this approach. Afterwards an effective boundary theory can be computed from which the Casimir energy follows from a one-dimension-less computation.

Here, we will analyze another intriguing possibility where such unexpected mass scale could originate from: the interactions induced by the boundary conditions. For this, we introduce a scalar toy model to mimic the non-linear YM boundary conditions. Indeed, due to gauge invariance, the boundary conditions on the plates must be expressed in terms of the non-Abelian field strength which possesses both a linear term in the gauge potential and a quadratic term coming from the commutator. Therefore, the proper implementation of the non-Abelian version of the common perfect electric/magnetic boundary conditions requires the analysis of boundary conditions which are non-linear in the gauge potential. The non-linear boundary conditions for the scalar field in the present manuscript have been exactly designed to shed preliminary novel light on the non-perturbative contributions to the Casimir energy arising from the non-Abelian nature of the gauge field.

The fact that interactions restricted to a lower-dimensional subspace can lead to interesting new physics can also be appreciated from works like \cite{Giombi:2019enr} or \cite{Andrei:2018die}. In our case, the effective boundary field method allows to take these non-linear boundary interactions into explicit account and explore their possible consequences for what concerns the Casimir energy. We will provide evidence that a non-perturbative ``boundary mass generation'' can occur, and, correspondingly, a non-perturbative correction to the Casimir energy/force (which may even change sign, depending on the size of the coupling constant).

\section{Non-linear Neumann-like boundary conditions}
As it has already been emphasised, eventually we wish to explore the possibility of novel, genuinely non-Abelian effects entering the Casimir energy of Yang-Mills theory. Even restricting to leading order, whilst the gauge field then satisfies a free field equation in the bulk, the boundary conditions on the plates possess both a term which is linear in the derivative of the gauge potential, as well as a quadratic one. To be more precise, e.g. the gauge invariant non-Abelian perfect magnetic boundary conditions read
\begin{equation}\label{conditionsYM}
  \left.n_\mu F_{\mu\nu}^a\right|_{z=\pm L/2}=\left. \p_z A_i^a - \p_i A_z^a - gf^{abc} A_z^b A_i^c\right|_{z=\pm L/2}=0,
\end{equation}
where the index $i$ refers to the $(t,x,y)$ coordinates. Therefore, one can appreciate that the Neumann boundary conditions, see \eqref{conditionsN} below, are mimicking the Yang-Mills case. The analysis, when extrapolated to the Yang-Mills case, will then strongly suggest the interesting possibility of the coexistence of two ground states (a trivial and a non-perturbative one) with
different energies but separated from one another via infinite energy walls. The underlying new physical effect is the generation of a non-perturbative boundary (gluon) mass.

Concretely, let us start from the following scalar toy model:
\begin{equation}\label{actie1}
S=\int d^4x\left[-\frac{1}{2} \varphi (\p^2-m^2)\varphi - b\delta\left(z+\frac L 2\right)\left(\p_z\varphi+\frac{g}{2}\varphi^2\right) -\bar b \delta\left(z-\frac L 2\right)\left(\p_z\varphi+\frac{g}{2}\varphi^2\right)\right]
\end{equation}
where the $b(\cev x)$ and $\bar b(\cev x)$ auxiliary fields impose the boundary conditions on the idealized (i.e.~zero thickness) plates,
\begin{equation}\label{conditionsN}
  \left.\p_z\varphi + \frac{g}{2}\varphi^2\right|_{z=\pm L/2}=0.
\end{equation}
We took the same $g$ on both plates for simplicity. We introduced the notation $\cev x = (t,x,y)$ for the $\mathbb R^3$ space orthogonal to the planes, and we are working in Euclidean space. The mass dimension of the fields or coupling $g$ is determined via
$\dim(b)=\dim(\bar b)=1$, $\dim(g)=0$.

Let us now look in detail at the quantum consequences of such non-linear boundary  conditions. The idea is to allow for a dynamically generated boundary mass, so we will construct the effective potential for $b_0$ and $\bar b_0$ with
\begin{eqnarray}\label{vevs}
   \left\{\begin{array}{ccccc}
            b & = & b_0 + \beta &;& b_0=\braket{b},\\
            \bar b & = & \bar b_0 + \bar\beta&;& \bar b_0=\braket{\bar b}.
          \end{array}\right.
   \end{eqnarray}
Let us first construct the two-point Green's function for the $\varphi$-field, but not yet taking into account the boundary conditions. The relevant quadratic part of the action reads
\begin{eqnarray}\label{action2}
S_2=\int d^4x\left[-\frac{1}{2} \varphi (\p^2-m^2)\varphi + \Delta(z)\varphi^2\right].
   \end{eqnarray}
We set $\Delta(z)=-\frac{g}{2}\left(b_0\delta_+ +\bar b_0 \delta_-\right)$ with $\delta_\pm\equiv \delta(z\pm L/2)$. Notice that this $\Delta(z)$ indeed acts as a boundary mass (per plate).

The Green's function $G(x,x')$ will thence obey
\begin{eqnarray}\label{green1}
\left(- \p^2+m^2+\Delta(z)\right)G(x,x')=\delta(x-x'),
   \end{eqnarray}
reducing to, after a partial $3D$ Fourier transformation
$\cev x \mapsto \cev k$ with $k=|\cev k|$,
\begin{eqnarray}\label{green2}
\left(- \p_z^2+k^2+m^2+\Delta(z)\right)G(\cev k, z,z')\equiv \mathcal M(z) G(\cev k, z,z')=\delta(z-z').
   \end{eqnarray}
For $\Delta(z)=0$, we have the standard free Green's function,
\begin{eqnarray}\label{green3}
G_0(\cev k, z,z')=\frac{1}{2\sqrt{k^2+m^2}}e^{-|z-z'|\sqrt{k^2+m^2}}.
   \end{eqnarray}
To avoid notational clutter, from here on we will omit the Fourier coordinate $\cev k$ from the Green's functions. Continuing, the complete Green's function is formally given by
\begin{eqnarray}\label{green4}
G(z,z')=G_0(z,z')-\int dx G_0(z,x)\Delta(x) G(x,z'),
   \end{eqnarray}
or, concretely
\begin{eqnarray}\label{green4b}
G(z,z')=G_0(z,z')+gb_0G_0(z,-L/2)G(-L/2,z')+g\bar b_0 G_0(z,L/2)G(L/2,z').
   \end{eqnarray}
From this equation, we can infer
\begin{eqnarray}\label{green4c}
\left\{\begin{array}{ccc}
         G(-L/2,z') & = & G_0(-L/2,z')+gb_0G_0(-L/2,-L/2)G(-L/2,z')+g\bar b_0G_0(-L/2,+L/2)G(+L/2,z') \\
         G(+L/2,z') & = & G_0(+L/2,z')+gb_0G_0(+L/2,-L/2)G(-L/2,z')+g\bar b_0G_0(+L/2,+L/2)G(+L/2,z')
       \end{array}
\right.
   \end{eqnarray}
and solve these for
\begin{eqnarray}\label{green4d}
\left\{\begin{array}{ccc}
         G(-L/2,z') & = & \displaystyle -\frac{g\bar b_0e^{-\sqrt{k^2+m^2}(L+|L/2-z'|)}+e^{-\sqrt{k^2+m^2}|L/2+z'|}(-g\bar b_0+2\sqrt{k^2+m^2})}{g^2b_0\bar b_0 e^{-2L\sqrt{k^2+m^2}}-(gb_0-2\sqrt{k^2+m^2})(g\bar b_0-2\sqrt{k^2+m^2})}  \\
        G(+L/2,z') & = & \displaystyle -\frac{g b_0e^{-\sqrt{k^2+m^2}(L+|L/2+z'|)}+e^{-\sqrt{k^2+m^2}|L/2-z'|}(-g b_0+2\sqrt{k^2+m^2})}{g^2b_0\bar b_0 e^{-2L\sqrt{k^2+m^2}}-(gb_0-2\sqrt{k^2+m^2})(g\bar b_0-2\sqrt{k^2+m^2})}
              \end{array}
\right.
   \end{eqnarray}
to end up with a closed analytical expression for $G(z,z')$ in \eqref{green4b}.

We are now armed to construct the effective potential. We start from
\begin{eqnarray}\label{V1}
  Z&=&\int [\mathcal{D}\varphi][\mathcal{D}\beta][\mathcal{D}\bar\beta] \exp\left[-S_2-\int d^4x \, \varphi\left((b_0+\beta)\delta_+ + (\bar b_0+\bar\beta)\delta_-\right)\right] \nonumber\\
  &=& \int [\mathcal{D}\beta][\mathcal{D}\bar\beta] \exp\left[\frac{1}{2}\int \frac{d^3k}{(2\pi)^3} \frac{d^3k'}{(2\pi)^3} dz dz' \left(J(\cev k,z) G(z,z')(2\pi)^3\delta^3(\cev k-\cev k') J(\cev k', z')\right)\right]\exp\left(-\frac{1}{2} \ln \det \mathcal M\right),
\end{eqnarray}
where the operator $\mathcal{M}(z)$ was defined in \eqref{green2}. $J(\cev k,z)$ is the $3D$ Fourier transform of the combination of terms in $Z$ coupling to $\varphi$, that is,
\begin{eqnarray}\label{V2bis}
J(\cev k, z)&=&-\left(b_0(2\pi)^3\delta^{(3)}(\cev k) + \beta(\cev k)\right)\delta_+^\prime - \left(\bar b_0(2\pi)^3\delta^{(3)}(\cev k) + \bar\eta(\cev k)\right)\delta_-^\prime.
\end{eqnarray}
First, we shall deal with the $\det \mathcal M$. Let us call $u_n(z)$ an eigenfunction of $\mathcal M$ with corresponding eigenvalue $\lambda_n$,
\begin{eqnarray}\label{schr1}
\mathcal M u_n=\lambda_n u_n.
   \end{eqnarray}
It will be sufficient to focus on the zero mode $\lambda_0=0$.
For the normalizable solution, we find
\begin{eqnarray}
  u_0(z) &=& \left\{\begin{array}{lcc}
               a e^{\sqrt{k^2+m^2}z} & \text{for} & z<-L/2 ,\\
               b e^{-\sqrt{k^2+m^2}z} + c e^{\sqrt{k^2+m^2}z}& \text{for} & -L/2<z<L/2 ,\\
               d e^{-\sqrt{k^2+m^2}z} & \text{for} & L/2<z.
             \end{array}\right.
\end{eqnarray}
The ``gluing'' at the boundaries $z=\pm L/2$ easily happens by imposing $u_0(z)$ to be continuous, whilst integrating \eqref{schr1} over either $\pm\epsilon-L/2$ or $\pm\epsilon+L/2$, respectively yielding
\begin{eqnarray}\label{bd1}
\left\{\begin{array}{rcc}
          u_0(-L/2-\epsilon)- u_0(-L/2+\epsilon)& = &0 ,\\
          u_0(+L/2+\epsilon)- u_0(+L/2-\epsilon) & = & 0 ,\\
          u_0'(-L/2+\epsilon)-u_0'(-L/2-\epsilon)+gb_0 u(-L/2) & = & 0 ,\\
          u_0'(+L/2+\epsilon)-u_0'(+L/2-\epsilon)+g\bar b_0 u(+L/2) & = & 0,
        \end{array}\right.
\end{eqnarray}
always assuming $\epsilon\to 0^+$. Denoting the matrix of the linear set of equations \eqref{bd1} by $\mathcal{N}$, it follows that $\det \mathcal{N}=0$ for a non-trivial solution to survive.

In fact, even if $\lambda_n\neq0$, a very similar reasoning can be made, leading to the requirement that normalizable eigensolutions will exist iff $\det\mathcal{N}(\lambda)=0$ for $\lambda=\lambda_n$. Said otherwise, we have found a function $\det\mathcal N(\lambda)$ (an ordinary rather than a functional determinant) that vanishes at the eigenvalues of the original problem \eqref{schr1}.

As such, we can invoke the Gel'fand-Yaglom approach \cite{gel1960integration} to extract $\ln\det \mathcal{M}$, following the line of reasoning in\footnote{A more detailed version can be accessed via \url{https://saalburg.aei.mpg.de/wp-content/uploads/sites/25/2017/03/dunne.pdf}, including more references.} \cite{Dunne:2007rt}. (An alternative approach would consist of using the identity of Jacobi for computing $\frac{d}{dL}\ln\det \mathcal{M}$.) In short, using $\zeta$-function regularization, one has
\begin{equation}\label{zeta1}
  \det\mathcal M = e^{-\zeta'(0)},\qquad \zeta(s)=\sum_n \frac{1}{\lambda_s^n}.
\end{equation}
It can then be shown that
\begin{equation}\label{zeta2}
  -\zeta'(0)= \ln \det \mathcal{N}(0) - \ln \det \mathcal{N}(-\infty).
\end{equation}
For $\lambda\to-\infty$, $\mathcal{M}(z)\to \mathcal{M}_{\text{free}}=-\partial_z^2$, so we can write
\begin{equation}\label{zeta3}
  \ln\det \mathcal{M}-\ln\det \mathcal{M}_{\text{free}}= \ln \det\mathcal{N}(0) -\ln \det \mathcal{N}_{\text{free}}(0).
\end{equation}
In our case, $\ln\det \mathcal{M}_{\text{free}}$ and $\ln \det\mathcal{N}_{\text{free}}(0)$ will merely be infinite constants that are irrelevant for the eventual Casimir energy/force, as an equal infinite constant will be subtracted when considering $E_\text{Cas, N}(L)-E_\text{Cas, N}(\infty)$. Or, alternatively, in e.g.~dimensional regularization, these infinite constants are replaced by zero, leading to the same end result.
So,
\begin{equation}\label{zeta4}
  \ln\det \mathcal{M}=\ln \det\mathcal{N}(0).
\end{equation}
After some manipulations, we find
\begin{equation}\label{zeta5}
  \ln\det \mathcal{M}= \ln\left(g^2b_0\bar b_0(1-e^{-2L\sqrt{k^2+m^2}})+4k^2+4m^2-2g(b_0+\bar b_0)\sqrt{k^2+m^2}\right).
\end{equation}
It then remains to compute
\begin{equation}\label{else1}
 \int [\mathcal{D}\beta][\mathcal{D}\bar\beta] \exp\left(\frac{1}{2}\int \frac{d^3k}{(2\pi)^3} \frac{d^3k'}{(2\pi)^3} dz dz' \left[J(\cev k,z) G(z,z')(2\pi)^3\delta^3(\cev k-\cev k') J(\cev k', z')\right]\right).
\end{equation}
The terms linear in $\beta$ or $\bar\beta$ will vanish upon proper minimization of the effective potential. Working out the various $\delta_\pm$'s, we get
\begin{equation}\label{else2}
  \eqref{else1}=\exp\left(\frac{V_3}{2}\lim_{\cev k\to0}\left(
                                   \begin{array}{cc}
                                                     b_0 & \bar b_0 \\
                                                   \end{array}
                                                 \right)
   \cdot \mathcal P \cdot \left(
                            \begin{array}{c}
                              b_0 \\
                              \bar b_0 \\
                            \end{array}
                          \right)
   \right) \int [\mathcal{D}\beta][\mathcal{D}\bar\beta]\exp\left(\frac{1}{2}\int\frac{d^3k}{(2\pi)^3}\left(
                                                   \begin{array}{cc}
                                                     \beta(\cev k) & \bar \beta(\cev k) \\
                                                   \end{array}
                                                 \right)
   \cdot \mathcal P \cdot \left(
                            \begin{array}{c}
                              \beta(-\cev k) \\
                              \bar \beta  (-\cev k) \\
                            \end{array}
                          \right)\right),
\end{equation}
where we defined
\begin{equation}\label{else3b}
  \mathcal{P}=\left(
                \begin{array}{cc}
                  \p_z\p_{z'}G(-L/2,-L/2) & \p_z\p_{z'}G(-L/2,+L/2) \\
                  \p_z\p_{z'}G(+L/2,-L/2) & \p_z\p_{z'}G(+L/2,+L/2) \\
                \end{array}
              \right).
\end{equation}
We also identified $V_3=(2\pi)^3\delta^{(3)}(\cev k=0)$ for the $3D$ space volume. We pause here to draw attention to the power of the auxiliary fields method to impose the boundary conditions. Although a priori these were merely multipliers, they do couple to the bulk dynamics and by integrating out the latter, one ends up with a quantum corrected boundary field sector, including non-trivial propagation and induced interactions purely living on the boundaries. We restrict ourselves to the leading order corrections here (Gaussian approximation) but in principle the $1PI$ effective action $\Gamma(\beta, \bar\beta)$ can be computed to higher orders.

Putting everything back together, we obtain
\begin{eqnarray}\label{V1bis}
  Z&=&\exp-\left(\frac{1}{2}\ln\det\mathcal M + \frac{1}{2}\ln\det(-\mathcal P)-\frac{V_3}{2}\lim_{\cev k\to0}\left(
                                                   \begin{array}{cc}
                                                     b_0 & \bar b_0 \\
                                                   \end{array}
                                                 \right)
   \cdot \mathcal P \cdot \left(
                            \begin{array}{c}
                              b_0 \\
                              \bar b_0 \\
                            \end{array}
                          \right)\right).
\end{eqnarray}
The first two contributions can be nicely combined, viz.
\begin{eqnarray}\label{V1tris}
  \frac{1}{2}\ln\det\mathcal M + \frac{1}{2}\ln\det(-\mathcal P)= \frac{1}{2}\ln\det(\mathcal M\cdot \mathcal P)=\frac{1}{2}\ln \det\left[(1-e^{-2L\sqrt{k^2+m^2}})(k^2+m^2)^2\right]
\end{eqnarray}
where we dropped an irrelevant infinite constant from switching $-\mathcal{P}$ with $\mathcal{P}$. Thus,
\begin{eqnarray}\label{V1bis}
  Z&=&\exp\left(\frac{V_3}{2}\lim_{\cev k\to0}\left(
                                                   \begin{array}{cc}
                                                     b_0 & \bar b_0 \\
                                                   \end{array}
                                                 \right)
   \cdot \mathcal P \cdot \left(
                            \begin{array}{c}
                              b_0 \\
                              \bar b_0 \\
                            \end{array}
                          \right)\right)\exp\left[ -\frac{1}{2}\text{tr}\ln \left((1-e^{-2L\sqrt{k^2+m^2}})(k^2+m^2)^2\right)\right]
                          ~=~ \exp(-V_3 E_\text{Cas, N}),
\end{eqnarray}
from which we can read off the Casimir energy in the Neumann case,
\begin{eqnarray}\label{V1bis}
    E_\text{Cas, N}&=&-\frac{1}{2}\lim_{\cev k\to0}
    \left(
        \begin{array}{cc}
            b_0 & \bar b_0 \\
        \end{array}
    \right)
    \cdot \mathcal P \cdot
    \left(
        \begin{array}{c}
            b_0 \\
            \bar b_0 \\
        \end{array}
    \right)
    +\frac{1}{2}\int \frac{d^3k}{(2\pi)^3}\left[(1-e^{-2L\sqrt{k^2+m^2}})(k^2+m^2)^2\right]
    \equiv E_\text{Cas, N}^{(1)}+E_\text{Cas, N}^{(2)}.
\end{eqnarray}
The first part is evidently only relevant in case of non-vanishing VEVs, so let us first focus on the second part, being
\begin{eqnarray}\label{evac1}
  E_\text{Cas, N}^{(2)}&=&\frac{1}{2}\int \frac{d^3k}{(2\pi)^3}\ln\left[(1-e^{-2L\sqrt{k^2+m^2}})(k^2+m^2)^2\right]~=~\int_0^\infty \frac{dk}{4\pi^2} k^2\ln(1-e^{-2L\sqrt{k^2+m^2}})~=~-\frac{m^2}{4\pi^2L}\sum_{n=1}^{+\infty}\frac{K_2(2nmL)}{n^2}
  \end{eqnarray}
up to an $L$-independent term that we dropped. It is reassuring that we recover the well-known fact that the standard Neumann Casimir energy equals that of the Dirichlet case, see e.g.~\cite{Ambjorn:1981xw}.
This expression \eqref{evac1} has a non-singular limit in the massless case, $\lim_{m\to0}E_\text{Cas, N}^{(1)}=\frac{-\pi^2}{1440L^3}$.

More interestingly, let us look at the new piece,
\begin{equation}\label{evac2N}
  E_\text{Cas, N}^{(1)}=  -\frac{gm^2b_0\bar b_0(b_0+\bar b_0)-m^2e^{2Lm}(-g b_0\bar b_0(b_0+\bar b_0)+2m(b_0^2+\bar b_0^2))-2mb_0\bar b_0e^{Lm}(g^2b_0\bar b_0-gm(b_0+\bar b_0)+2m^2)}{2 \left(e^{2 L m} (gb_0 -2 m) (g\bar b_0-2 m)-g^2b_0\bar b_0\right)}.
  \end{equation}
Closed expressions exist for the various (complex) extrema of $E_\text{Cas, N}^{(1)}$ in terms of $(b_0,\bar b_0)$, but these are lengthy algebraic expressions that are not very illuminating. We will therefore present results for the non-zero mass case in a purely numerical fashion.

\begin{figure}[h]
  \centering
    \includegraphics[width=0.4\textwidth]{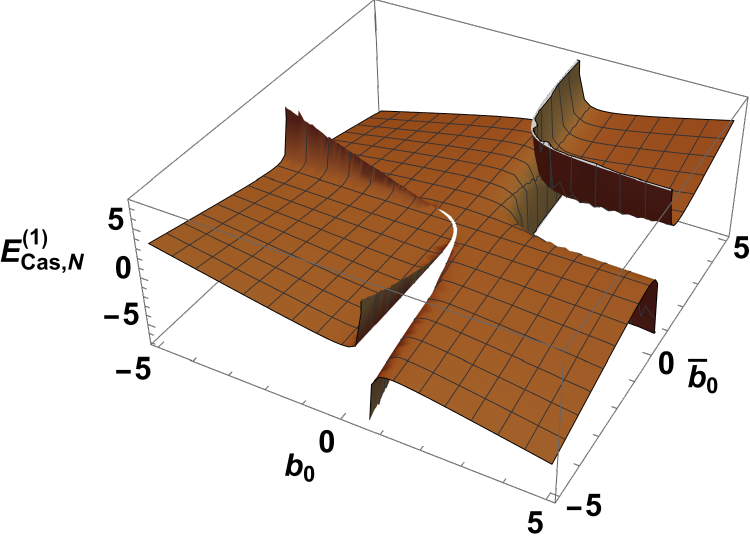} \qquad\qquad
    \includegraphics[width=0.4\textwidth]{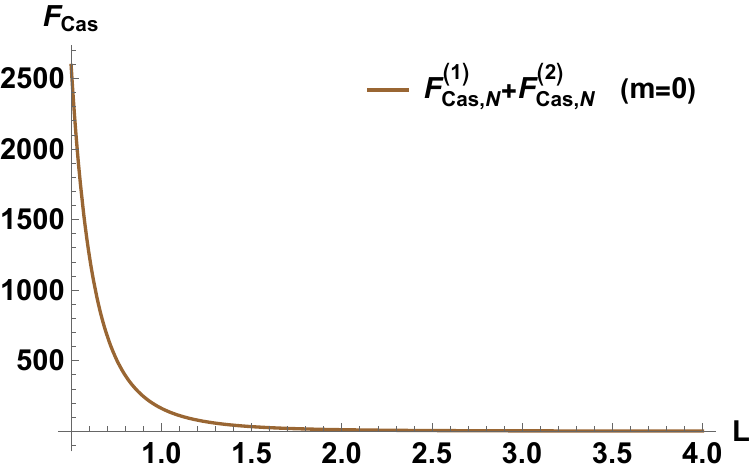}
  \caption{(Left) $E_\text{Cas, N}^{(1)}$ from \eqref{evacmassless0} for $g=\frac{1}{2}$, $L=1$.; (Right) The corresponding non-linear Neumann Casimir force.}\label{CasFN}
\end{figure}

However, let us first focus on the $m=0$ case, as this case comes closest to the YM one. Everything can be handled fully analytically then. Indeed, as
\begin{eqnarray}\label{evacmassless0}
  E_\text{Cas, N}^{(1)}&=&  \frac{g b_0^2 \bar b_0^2}{2gLb_0\bar b_0-2(b_0+\bar b_0)},
  \end{eqnarray}
we find a non-trivial minimum at $b_0=\bar b_0 = \frac{3}{gL}$ for all $L$, with Hessian eigenvalues $\left\{\frac{3}{2L}, \frac{9}{2 L}\right\}$. There is also the trivial minimum at $b_0=\bar b_0=0$, with lower energy. However, both are genuine groundstates as one cannot penetrate (tunnel through) the infinite energy barriers separating both sectors, see FIG.~\ref{CasFN}. Likewise, even for large fields, there is no way to tunnel to the regions where the potential drops to $-\infty$ because of the energy walls.

In the non-trivial groundstate, in total we end up with
\begin{eqnarray}\label{evacmassless}
  E_\text{Cas, N}&=& E_\text{Cas, N}^{(1)}+E_\text{Cas, N}^{(2)}= \frac{27}{2g^2L^3} -\frac{\pi^2}{1440L^3}\Longrightarrow F_\text{Cas, N}=-\p_L E_\text{Cas, N}=\frac{81}{2g^2L^4}-\frac{\pi^2}{480L^4}.
  \end{eqnarray}
  Interestingly, for small values of $g$ the Casimir force is repulsive, while there is a critical $g_c= \frac{36\sqrt{15}}{\pi}\approx 44.4$  at which the Casimir force changes sign.

\vspace{\baselineskip}
For the massive case,  we will again choose $g=\frac{1}{2}$ and work in units $m=1$, the behaviour as reported below is generic for other parameter choices. We will also only focus on the real minima, as there are several other (complex) saddle points or local maxima of no interest to us. For $L<1.6$, the trivial solution remains an absolute minimum, whilst (again shielded with an infinite energy barrier from the trivial minimum or unstable regions) another local minimum appears at $b_0=\bar b_0\approx 8.3$.

For larger $L$, $L>1.6$, $(0,0)$ remains a global minimum, while the other minimum evolves into a saddle point or local maximum if $L$ grows further. Overall, $(0,0)$ is thus a global minimum in a region that is shielded with infinite energy barriers from unstable areas.

\section{Non-linear Dirichlet boundary conditions}
Let us also briefly discuss the non-linear Dirichlet case. The analysis is largely similar to the Neumann case, by simply redefining  $J(\cev k, z)$ as follows
\begin{eqnarray}\label{V2}
J(\cev k, z)&=&\left(b_0(2\pi)^3\delta^{(3)}(\cev k) + \beta(\cev k)\right)\delta_+ + \left(\bar b_0(2\pi)^3\delta^{(3)}(\cev k) + \bar\eta(\cev k)\right)\delta_-
   \end{eqnarray}
which leads to a redefinition of the matrix $\mathcal P$,
\begin{equation}\label{else3}
  \mathcal{P}=\left(
                \begin{array}{cc}
                  G(-L/2,-L/2) & G(-L/2,+L/2) \\
                  G(+L/2,-L/2) & G(+L/2,+L/2) \\
                \end{array}
              \right).
\end{equation}
Eventually, we find
\begin{eqnarray}\label{evac2}
  E_\text{Cas, D}^{(1)}&=&
  -\frac{e^{2 L m} \left(2 m \left(b_0^2+\bar b_0^2\right)-b_0 \bar b_0 g (b_0+\bar b_0)\right)+b_0 \bar b_0 g (b_0+\bar b_0)+4 b_0 \bar b_0 m e^{L m}}{2 \left(e^{2 L m} (gb_0 -2 m) (g\bar b_0-2 m)-g^2b_0\bar b_0\right)}.
\end{eqnarray}
For $m\neq0$, there is a non-trivial extremum at
\begin{equation}\label{sol1}
  b_0=\bar b_0 = \frac{4m e^{Lm}}{g(1+e^{Lm})}.
\end{equation}
The Hessian eigenvalues associated to \eqref{sol1} are given by $\left\{\frac{1+e^{-Lm}}{2m},\frac{\sinh(Lm)}{m(-3+e^{Lm})}\right\}$. As such, \eqref{sol1} corresponds to a non-perturbative local minimum iff
\begin{equation}\label{sol3}
  L>\frac{\ln 3}{m} \;,
\end{equation}
and a saddle point otherwise.

As such, in the Dirichlet case, new dynamics emerges only in the massive case, and for sufficiently large plate separation. Said otherwise, we have found a non-trivial stable absolute minimum, shielded from tunneling to other regions because of infinite energy walls. We note here that expression \eqref{evac2}, just as its Neumann counterpart \eqref{evac2N}, does not explicitly depend on the space-time dimensionality, which can be easily seen from the expression \eqref{else2} for what concerns the $\lim_{\cev k\to0}\left(
                                                   \begin{array}{cc}
                                                     b_0 & \bar b_0 \\
                                                   \end{array}
                                                 \right)
   \cdot \mathcal P \cdot \left(
                            \begin{array}{c}
                              b_0 \\
                              \bar b_0 \\
                            \end{array}
                          \right)$ contribution.

For completeness, we mention there is also a trivial maximum at
\begin{equation}\label{sol2}
  b_0=\bar b_0 = 0.
\end{equation}
The Hessian eigenvalues at $(0,0)$ are given by $\frac{-1\pm e^{-Lm}}{2m}$, so the trivial vacuum is unstable now. This is quite distinctive from the Neumann findings.
Furthermore, there are also two saddle points at $(b_0,\bar b_0)=(\bar b_0,b_0) = \left(\frac{4me^{2Lm}}{g(-1+e^{Lm})},\frac{4me^{Lm}}{g(-1+e^{Lm})}\right)$. It can be verified that the Hessian eigenvalues always have opposite signs in both cases.

For $L<\frac{\ln 3}{m}$, there is nowhere a minimum to be found, only the local maximum and three (complex) saddle points. Higher order corrections  to the Casimir energy in the coupling $g$ are necessary to decide whether a minimum reappears.  It would be most interesting to study this possible phase transition in terms of $L$ in a truly dynamical setting by considering a (slowly) moving plate, but the dynamical Casimir effect, see e.g.~\cite{Bordag:2001qi} and references therein, is another story and beyond the scope of this paper. Returning to the novel minimum \eqref{sol1}, under the condition \eqref{sol3}, we eventually end up with a Casimir energy given by
\begin{eqnarray}\label{eindcas}
    E_\text{Cas, D}&=& \frac{e^{Lm}}{1+e^{Lm}}\frac{m}{8g^2}   -\frac{m^2}{4\pi^2L}\sum_{n=1}^{+\infty}\frac{K_2(2nmL)}{n^2},
\end{eqnarray}
where we note that for large separations $L$, the non-perturbative first piece will dominate.\footnote{Strictly speaking, if we adopt the convention to define the Casimir energy subtraction so that $E_\text{Cas, D}(\infty)=0$, we should subtract $\frac{m}{8g^2}$ from \eqref{eindcas}. This is anyhow irrelevant for the Casimir force itself.} The Casimir force itself remains negative and thus repulsive in the Dirichlet case, even upon including the non-perturbative boundary mass effect, In the region $L<\frac{\ln 3}{m}$, we cannot make definite statements.

\section{Conclusions}
We implemented a set of non-linear boundary conditions for an otherwise free scalar field theory by means of auxiliary fields restricted to a lower-dimensional interface, in our case two infinite parallel plates. This problem is extremely relevant in itself since non-linear boundary conditions arise in many contexts (such as chemical diffusion \cite{KASHANI2022116925}, non-linear optics \cite{Agarwal:87}, vibration problems in nuclear reactor dynamics \cite{painter2022nonlinear}, etc.). As such, interactions are included on these interfaces, and we investigated the possibility that these can lead to a dynamical boundary mass generation, which consequently affects the Casimir energy in a non-perturbative manner.
We have given evidence, at least at leading order (mean field), that such is the case, both in the non-linear Dirichlet and Neumann models. Although the computations are quite similar, the results are rather different. The most interesting finding is that in the Neumann case with massless scalars, two ground states are found with different energies (a trivial and a non-perturbative one), but separated from one another via infinite energy walls. In the non-perturbative ground state, the Casimir force can either be repulsive or attractive depending on the strength of the coupling. In the Dirichlet case, only for massive scalars, and for sufficiently large enough plate separation, a new non-perturbative ground state is found, still leading to an attractive Casimir force.

On the other hand, besides the intrinsic interest in analyzing the effect of non-linear boundary conditions on the Casimir energy, the most important motivation stems from the fact that, in the Yang-Mills case, the field strength is quadratic in the gauge potential. In contrast, the boundary conditions for the Yang-Mills Casimir effect are linear in the field strength (and, therefore, quadratic in the gauge field itself). Thus, our findings strongly motivate the investigation of whether similar non-perturbative boundary effects can emanate from the gauge-invariant YM boundary conditions, which are automatically non-linear when written in terms of the field strength tensor. Hence, our results shed novel light on the recent series of lattice YM Casimir works \cite{Chernodub:2018pmt,Chernodub:2023dok,Ngwenya:2025mpo,Ngwenya:2025cuw}, which suggest the existence of a new light mass scale. We also expect an interesting interplay with the dynamical “bulk” gluon mass generation. In particular, the case of non-linear Neumann boundary conditions analyzed in the present manuscript suggests the possibility of coexisting vacua (one trivial and the other non-perturbative), which can significantly affect the Yang-Mills Casimir energy. We will revisit this interesting issue in future work.

\section*{Acknowledgments}
F.C.~has been funded by FONDECYT Grants No. 1240048, 1240043 and 1240247 and also by Grant ANID EXPLORACIÓN 13250014.
The work of D.D.~and T.O.~was supported by KU Leuven IF project C14/21/087.
P.P.~gladly acknowledges support from Charles University Research Center (UNCE 24/SCI/016).
L.R.~acknowledges the Ministero dell’Universit\`a e della Ricerca (MUR), PRIN2022 program (Grant PANTHEON 2022E2J4RK) for partial support.
The work of S.S.~was funded by FWO PhD-fellowship fundamental research (file number: 1132823N).
D.D., L.R.~and S.S.~are grateful for the hospitality at CECs and USS, where this work was finalized.

\bibliographystyle{apsrev4-2}
\bibliography{bibfile}

\end{document}